# A model for exciton-polaritons in uniaxial molecular crystals describing spatial dispersion, refraction and reflection

Stefan C.J. Meskers[1*] and Girish Lakhwani[2]

[1] *Molecular Materials and Nanosystems and Institute for Complex Molecular Systems, Eindhoven University of Technology, NL-5600 MB Eindhoven, Netherlands.*

[2] *School of Chemistry, The University of Sydney, NSW 2006 Australia.*

Propagation of light through a uniaxial material is studied using field theoretical methods. The materials is modeled by cubic lattice of oriented classical Lorentz oscillators. A two-step coarse graining approach is applied. At the bulk level, excitations of the coupled light-matter system, or polaritons, are described by a Proca-type equation for massive vector bosons. On the microscopic level, multiple scattering is used to relate the sub-luminal speed of the polaritons to the polarizability of the Lorentz oscillators. For each direction of propagation of the polaritons, *three* independent polarizations exist, consistent with the integer spin of massive vector bosons. Reflection and refraction are calculated by imposing the requirement of a uniform gauge for the electromagnetic vector potential across the interface of the uniaxial molecular material and vacuum. Reflectance spectra near the resonance frequency are calculated. The spectra feature a characteristic minimum in middle of the reflection band, in agreement with experiment. An incident unpolarized light beam is predicted to refract into *three* different rays. The model supports surface bound excitations and predicts a Goos-Hänchen shift of the reflected beam upon reflection of light incident from vacuum onto the material.

* s.c.j.meskers@tue.nl

## Popular Summary

Inside a material, a photon hybridizes with electronic excitations to form an exciton-polariton. The exciton-polariton has three available spin states or polarizations, while a photon in vacuum has only two. This change in the number of internal degrees of freedom is related to breaking of the gauge symmetry for the photon when going into polarizable matter where the speed of propagation of the energy quantum is reduced to below $c$. We argue that when calculating reflection and refraction of light at the vacuum/matter interface, a uniform gauge condition should be applied across the interface. This condition allows for a solution of the infamous additional boundary condition (ABC) problem in the optics of condensed matter. We show that experimental reflection spectra from molecular crystals can be reproduced accurately. In addition, we predict a lateral displacement of a light beam reflected externally at the vacuum/matter interface with respect to the incoming beam, an effect known as the Goos-Hänchen shift. Furthermore, the model predicts the occurrence of *trirefringence*, i.e. the refraction of an incident light beam into three separate beams with different polarization when entering an anisotropic material.

# I. INTRODUCTION

Inside matter, electromagnetic energy is known to occur in quanta called polaritons. In the frequency range where electronic excitation of the material can take place, one speaks of exciton-polaritons. An exciton-polariton can be thought of as a quasi-particle that is a hybrid between a photon and an electronic excitation of the solid. Following the seminal work of Hopfield and Pekar [1-3], research on polaritons has flourished [4,5] with important recent developments being cavity polaritons [6], polariton lasing [7,8] and polariton condensation [9-11]. Furthermore, advances in organic electronics have heightened the prospects of electrically driven polariton lasing in organic diodes [12-14]. Low threshold polariton lasing involves Bose-Einstein condensation of polaritons in the organic material [15] and, obviously, engineering of this optoelectronic property requires detailed understanding of the polaritons modes in the organic layer.

A problem in the description of the optical properties of matter involving polaritons is that additional boundary conditions need to be introduced in order to describe the behavior of the polaritons near the interface of the material [16]. For calculation of reflection and refraction, the boundary conditions for the electrical polarization associated with the polaritons are of prime importance, Yet, these boundary conditions are still a matter of debate [17-25]. For inorganic semiconductors, the problem of the addition boundary conditions could recently be solved by a microscopic approach in which the quantum mechanical motion of the electrons under the influence of the electromagnetic fields is evaluated directly, avoiding the introduction of a macroscopic polarization function [22,24]. For a molecular material consisting of organic dye molecules such a direct quantum mechanical approach is hardly feasible because of the computational difficulties associated with the strong electron-electron correlation in organic molecules and the more localized nature of electron states.

In this contribution, we resolve the issue of the additional boundary conditions for polaritons in molecular crystals. The route we follow is to formulate a Lagrangian for the coupled field-matter system that is invariant under Lorentz transformation and involves the vector potential *A* for the electromagnetic degrees of freedom. This field-theoretical approach gives full inside into the gauge freedom of the problem of reflection of light. Boundary conditions for reflection and refraction at the interface between vacuum and the dipole lattice are then be derived, taking into account the additional requirement of a uniform gauge condition across the vacuum-matter interface. This allows for a comprehensive description of optical effects associated with reflection and refraction. The model predicts reflection spectra, the refraction of an unpolarized light beam into three separate rays ('trirefringence'), the existence surface bound waves and the possibility of a Goos-Hänchen shift of the reflected beam for light incident from vacuum onto the material's surface. The results are relevant for a detailed understanding of the in-coupling and out-coupling of light into/out of organic semiconductors in for instance organic photovoltaic cells [26] and organic light emitting diodes [27]. Furthermore, the insight into the propagation of electromagnetic waves in the molecular solid should also enables a deeper understanding of the 'diffusion length' of electronic excitations in these molecular solids [28].

The organization of the paper is as follows. In subsection II.A, we introduce a model for the uniaxial material, comprising a cubic lattice of oriented dipole oscillators. For this model we then derive a Lorentz covariant Lagrange density for the coupled mechanical and electromagnetic degrees of freedom. In II.B we discuss how to gauge the vector potential for the polaritons. In II.C we derive the possible polarization states for the polaritons in the bulk uniaxial medium. Then in II.D we switch from a macroscopic to a microscopic view and discuss the polarizability of the individual oscillators. In II.E & F we then relate the speed of propagation of the polariton in the bulk to the microscopic polarizability of the individual oscillators on the lattice points. Combining the macroscopic and microscopic pictures allows for calculation of the dispersion relations of the polaritons, as presented in II.G. In the remainder of the manuscript we discuss application of the dispersion relations to the problem of reflection and refraction. In subsection II.H we derive the boundary conditions for the vector potential at the interface between vacuum and the cubic lattice of oscillators. Using these boundary conditions and the dispersion relation for the polaritons, we then calculate reflection spectra, see II.I. In II.J we investigate the refracted rays and discuss the prediction of *trirefringence* in the uniaxial medium. In II.K we combine the insights in reflection and refraction and discuss the prediction of surface bound waves and the associated Goos-Hänchen shift of beams reflected at the surface of the dipole lattice. Finally, in II.L we compare the predictions from the model to experimental reflectance data from literature.

## II. RESULTS

### A. Lagrange density for a cubic lattice of oriented dipole oscillators interacting with electromagnetic fields

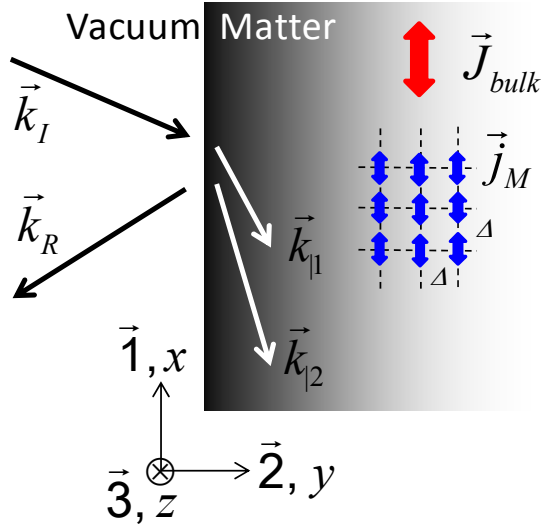

**FIGURE 1**. Geometry for reflection and refraction from a uniaxial material modeled by a cubic lattice of oriented dipole oscillators. The interface between vacuum and matter is described by $y = 0$. Light incident with wavevector $k_I$ in the $xy$ plane excites harmonic oscillators that are positioned on a cubic lattice with lattice constant $\Delta$ and are oriented in the 1 or $x$ direction. The oscillating current densities $j_M$ on the individual sites can be averaged to a bulk current density $J_{bulk}$.

The model we consider is a cubic lattice of one-dimensional, charged harmonic oscillators, oriented in the 1 or $x$-direction. The oscillators consist of a moving charge $q$ with mass $m_e$, harmonically bound to a static counter charge at the lattice point. We take the limit in which the lattice constant, $\Delta$, is small compared to the wavelengths of radiation of interest. In this limit, a continuous, mechanical displacement field $\eta$ can be introduced to describe the amplitude and phase of the oscillators. The Lagrangian for the square lattice of mechanical oscillators can then be expressed as [29,30]:

$$\mathcal{L}_{osc} = -\left[\mu(\partial_1\eta)^2 - Y(\partial_0\eta)^2\right] \tag{1}$$

where $\mu$ is the mass density ($m_e/\Delta^3$) and $Y$ is the Young's modulus per unit area. Mass density and Young's modulus determine the speed $v$ at which the longitudinal mechanical oscillations can propagate in the 1-direction: $v^2 = Y/\mu$.

In order to make the connection with electromagnetic fields we note that the first derivative in the 1 direction is related to the charge density in the follows way:

$$\partial_1\eta = \frac{\partial}{\partial x}\eta = \frac{\Delta^3}{q}\rho = \frac{\Delta^3}{q}J_0, \tag{2}$$

where the charge density $\rho$ has been expressed as the time-like component of the 4-current $J_\alpha$. Similarly, the derivative of the displacement field with respect to time gives the current density:

$$\partial_0\eta = \frac{\partial}{\partial t}\eta = \dot{x} = \frac{\Delta^3}{q}J_1 \tag{3}$$

Using expressions (2) and (3), the Lagrangian density of the lattice of oscillators can be put in manifestly covariant form:

$$\mathcal{L}_{osc} = -J_\alpha M^\alpha{}_\beta J^\beta \tag{4}$$

The complete Lagrangian for the lattice of charged mechanical oscillators together with the electromagnetic radiation [31] can now be given as:

$$\mathcal{L} = -\frac{1}{4}F^{\alpha\beta}F_{\alpha\beta} + J_\alpha A^\alpha - J_\alpha M^\alpha{}_\beta J^\beta \tag{5}$$

Where $F$ is the antisymmetric field tensor derived from exterior differentiation of the field amplitude of the four vector potential $A$: $F_{\mu\nu} = \partial_\mu A_\nu - \partial_\nu A_\mu$.

In order to get the equations of motion for the coupled system of matter and field, we apply the Euler- Lagrange relation to the Lagrange density (5). We first focus on the Euler-Lagrange equation for the field variables of the Lagrangian:

$$\partial_\beta \frac{\partial \mathcal{L}}{\partial(\partial_\beta A_\alpha)} - \frac{\partial \mathcal{L}}{\partial A_\alpha} = 0 \tag{6}$$

Applied to (5) and adopting the Lorentz gauge this gives:

$$\Box A^\alpha = -J^\alpha$$
$$\partial_\alpha A^\alpha = 0 \tag{7}$$

Next we consider the Euler-Lagrange equation with respect to the mechanical displacement field $\eta$:

$$\partial_0 \frac{\partial \mathcal{L}}{\partial(\partial_0 \eta)} + \partial_1 \frac{\partial \mathcal{L}}{\partial(\partial_1 \eta)} - \frac{\partial \mathcal{L}}{\partial \eta} = 0 \tag{8}$$

This gives

$$\partial_0 J_1 - \frac{Y}{\mu}\partial_1 J_0 = \omega_P^2 \left(\partial_0 A_1 - \partial_1 A_0\right) \tag{9}$$

with $\omega_P^2 = q^2 / \Delta^3 m$ the square of the plasma frequency.

We now assume that the electromagnetic radiation and the mechanic motion are in equilibrium and exhibit stable, cyclic motion. The actual frequency of the mechanical motion $\omega$ is locked to the frequency of the electromagnetic vector potential:

$$A_\alpha \propto e^{i\omega t - ik_1 x_1 - ik_2 x_2 - ik_3 x_3} \tag{10a}$$

$$J_1 \propto e^{i\omega t - ik_1 x_1} \tag{10b}$$

The mechanical oscillators have a natural bulk resonance frequency $\omega_Y$ that is related to the mass density and Young's modulus. We have:

$$\frac{Y}{\mu} = v^2 = \frac{\omega_Y^2}{k_1^2} \tag{11}$$

We can now rewrite (9) to

$$i\omega\left(1-\frac{\omega_Y^2}{\omega^2}\right)J_1 = \omega_P^2\left(i\omega A_1 - ik_1 A_0\right) \qquad (12)$$

Where we have made use of charge conservation $\partial_\alpha J^\alpha = 0$, implying for our case $k_1 J_1 = \omega J_0$

This results in:

$$J_1 = \omega_P^2 \frac{\omega^2 A_1 - k_1 \omega A_0}{\omega^2 - \omega_Y^2} \qquad (13)$$

By combing (13) with (7) we obtain a complete, macroscopic description for the coupled matter – field system.

### B. Lagragian for massive vector bosons and gauge freedom

Before we continue examining reflection and refraction, we need to take a more fundamental look at the allowed modes in the uniaxial material represented by the lattice of oriented dipole oscillators. We start by examining the Lagrangian for a particle with non-zero rest mass $\mu$ and spin one, i.e. a massive vector boson. The Lagrangian for such a particle was first studied by Proca:

$$\mathcal{L} = -\frac{1}{4}F^{\alpha\beta}F_{\alpha\beta} + \frac{\mu^2}{2}A_\alpha A^\alpha + J_\alpha A^\alpha \qquad (14)$$

From this Lagrangian, the following wave equation in reciprocal space can be derived when adopting the Lorentz gauge:

$$\begin{cases} -k_\beta k^\beta A^\alpha = \mu^2 A^\alpha - J^\alpha \\ k_\alpha A^\alpha = 0 \end{cases} \qquad (15)$$

The particle in vacuum without additional sources ($J = 0$) is fully characterized by *three* independent components of the four vector amplitude $A^\alpha$. The fourth component is fixed by the gauge condition. The existence of three independent components of the four vector potential is consistent with the number of degrees of freedom expected for a massive boson with spin quantum number $S = 1$ and the associated three possible values for the 'magnetic' quantum number $m_S = 0, \pm 1$. The field associated with a massive vector boson can have a longitudinal component, *viz*. if the particle moves in the *x*-direction then the component $A_x$ can be nonzero. In contrast, for massless vector bosons such as photons, longitudinal components are usually set to zero.

Because of the prime importance of the possible polarizations of the polariton in the description of reflection and refraction, we review below in detail under which conditions longitudinal components for vector bosons may be set to zero. We recall the wave equation in *k*-space for the photon in the Lorentz gauge:

$$\begin{cases} -k_\beta k^\beta A^\alpha = -J^\alpha \\ k_\alpha A^\alpha = 0 \end{cases} \quad (16)$$

Importantly, in the wave equation for massless bosons (16), there is additional remaining gauge freedom. In (16) the vector potential is not fully specified and replacing *A* by *A'* with $A'_\nu \to A_\nu - ik_\nu \chi$, we obtain exactly the same wave equation provided that the arbitrary function of space-time coordinates $\chi$ satisfies $\Box\chi=0$ or alternatively $k_\alpha k^\alpha \chi(k)=0$. This additional gauge freedom does not exist for the massive vector bosons. To see what this additional gauge freedom implies we imagine a massless vector boson travelling along the *y*-direction. The wave equation then demands $k_y = \omega$. A general solution to the wave equation obeying the Lorentz gauge condition is:

$$A^\alpha = \left(e^{i\omega t - ik_y y}\right) \begin{bmatrix} \Phi \\ A_x \\ \Phi \\ A_z \end{bmatrix} \quad (17)$$

However this solution is not unique and may be replaced by $A'_\alpha \to A_\alpha + \partial_x \chi$:

$$A'^{\alpha} = \left(e^{i\omega t - ik_y y}\right)\begin{bmatrix} \Phi \\ A_x \\ \Phi \\ A_z \end{bmatrix} + k_{\chi}^{\alpha} \chi \qquad (18)$$

with $\chi$ an arbitrary harmonic function, satisfying $\Box \chi = 0$. We can choose the spatial part of the wave vector $k_\chi$ of $\chi$ in the same direction as the wavevector of the photon $k_y$. Harmonicity then demands $k_\chi = k_y = \omega$. Furthermore we can choose $\chi = -\Phi/\omega$ such that we derive at:

$$A'^{\alpha} = \left(e^{i\omega t - ik_y y}\right)\left(\begin{bmatrix} \Phi \\ A_x \\ \Phi \\ A_z \end{bmatrix} - \begin{bmatrix} \Phi \\ 0 \\ \Phi \\ 0 \end{bmatrix}\right) = \left(e^{i\omega t - ik_y y}\right)\begin{bmatrix} 0 \\ A_x \\ 0 \\ A_z \end{bmatrix} \qquad (19)$$

This shows that longitudinal, $y$-component of the vector amplitude for the massless boson can be 'gauged' out. As a well know result, it follows that massless bosons have only two independent components or polarizations. Conversely, for a massive vector boson with $|\vec{k}| \neq \omega$, the additional gauge symmetry that exists for the massless case is broken, and the longitudinal component of the field cannot be categorically set to zero.

Finally we refer back to the Lagrangian (5) which we derived for the coupled electromagnetic and mechanical degrees of freedom for the lattice of oriented dipole oscillators oriental. As will be discussed in more detail below, this Lagrangian contains a term expressed in Eq. 4 that can be interpreted as resulting from an *anisotropic* mass. The anisotropic mass term breaks the additional gauge symmetry that exists for the photon in vacuum, except for specific directions of propagation and polarization.

### C. A uniaxial polarizbale medium and its polariton modes

In this section we examine the coupled mechanical and electromagnetic excitations in the bulk of the cubic lattice of oriented dipole oscillators in greater detail. We assume a finite polarizability of the bulk in the direction of the oscillators ($x$), characterized in a phenomenological way by the parameter $\alpha_B$. Using this bulk polarizability, we can relate the locally averaged, induced current density to the amplitude of the vector potential. In order to keep full track of the gauge invariance we use a covariant formulation. This gives then [32]:

$$J^\alpha = P^\alpha_\beta A^\beta \tag{20}$$

where the tensor $P$ describes the response of the material. Conditions that have to be fulfilled by $P$ are the conservation of charge

$$\partial_\alpha J^\alpha = 0 \Rightarrow k_\mu P^\mu_\nu = 0 \tag{21}$$

and the Lorentz gauge condition

$$\partial_\alpha A^\alpha = 0 \Rightarrow P^\alpha_\beta k^\beta = 0 \tag{22}$$

Based on these conditions we can then give a matrix representation of $P$

$$P^\mu_\nu = \begin{bmatrix} -\dfrac{k_1 k_1}{\omega^2}\alpha_B & \dfrac{k_1}{\omega}\alpha_B & 0 & 0 \\ -\dfrac{k_1}{\omega}\alpha_B & \alpha_B & 0 & 0 \\ 0 & 0 & 0 & 0 \\ 0 & 0 & 0 & 0 \end{bmatrix} \tag{23}$$

We note that the expression (23) for the polarizability is consistent with the earlier result (13), and establishes a relation between the phenomenological bulk parameters $\alpha_B$ and $\omega_Y$:

$$J_1 = P^\beta_1 A_\beta = \alpha_B \left( -\frac{k_1}{\omega} A_0 + A_1 \right) = \frac{\omega_P^2 \omega^2}{\omega^2 - \omega_Y^2}\left( A_1 - \frac{k_1}{\omega} A_0 \right) \tag{24}$$

**TABLE I**. Polariton modes in a polarizable uniaxial material. Non-zero components of the four vector potential $A = (A_0, A_1, A_2, A_3)$ for different directions and magnitude of the four wavevector $k$ are given. The material is taken to be polarizable only in the 1-direction.

|  | $k_\alpha k^\alpha \neq 0$ 'heavy' | $k_\alpha k^\alpha = 0$ 'light' |
|---|---|---|
|  | $A_1 \neq 0$ | $A_1 = 0$ |
| $k_1 = 0$ | $(0, A_1, 0, 0)$ <br><br> **one** indep. pol. transversal | $\left( \dfrac{k_2 A_2 + k_3 A_3}{\omega}, 0, A_2, A_3 \right)$ <br><br> **two** indep. pols partially longitudinal |
| $k_1 \neq 0$ | $(A_0, A_1, 0, 0)$ <br><br> **two** indep. pols partially longitudinal | $\left( 0, 0, -\dfrac{k_3}{k_2} A_3, A_3 \right)$ <br><br> **one** indep. pol. transversal |

We can now use (20) and (23) to reformulate the wave equation $\Box A^\alpha = -J^\alpha$ for the coupled matter radiation field in matrix form. Defining $n$ as the refractive index for the $x$-direction $n = k_1/\omega$ and $kk = k_\alpha k^\alpha = \omega^2 - k_1^2 - k_2^2 - k_3^2$ and adopting the Lorentz gauge we get:

$$\begin{bmatrix} kk - n^2 \alpha_B & \alpha_B n & 0 & 0 \\ -n \alpha_B & kk + \alpha_B & 0 & 0 \\ 0 & 0 & kk & 0 \\ 0 & 0 & 0 & kk \end{bmatrix} \begin{bmatrix} nA_1 + \dfrac{k_2}{\omega} A_2 + \dfrac{k_3}{\omega} A_3 \\ A_1 \\ A_2 \\ A_3 \end{bmatrix} = 0 \quad (25)$$

From Eq. 25 we can now derive the allowed modes inside the material and establish relations between the polarization of the mode and its dispersion relation, see Table I. A useful distinction is between 'heavy' modes with non-zero polariton mass, i.e. $k_\alpha k^\alpha \neq 0$, and 'light' modes with zero mass, $k_\alpha k^\alpha = 0$. From (25) it follows immediately that heavy modes must have $A_2 = A_3 = 0$. Conversely, modes with nonzero $A_2$ or $A_3$ have to obey the vacuum dispersion relation with $k_\alpha k^\alpha = 0$. Interestingly if we consider a heavy mode travelling in a direction perpendicular to the $x$-axis ($k_1=0$), it follows that the polariton can have only one polarization, namely in the 1-direction, with the only non-zero component of the four potential being $A_1$. The polarization of this mode is thus transversal ($\vec{k} \cdot \vec{A} = 0$). In addition, for the same direction of propagation, there are two independent 'light' modes with polarizations along $y$ and $z$, described by the independent components $A_2$ and $A_3$. These modes can have a longitudinal character. To be more precise, one of these two modes may be chosen to be fully transversal and the other longitudinal.

For modes traveling in a direction that has a component along $x$ ($k_1 \neq 0$), the mode structure is different. The 'heavy' modes are characterized by two independent components ($A_0$ and $A_1$). In Addition for these propagation directions there is one light mode with one independent component, which we take arbitrarily to be $A_3$. In summary, polaritons arising from coupled mechanical and electromagnetic degrees of freedom are characterized by three possible polarizations for all frequencies $\omega$ and directions $\vec{u} = \vec{k} / |\vec{k}|$ of the spatial part of the wave vector. Obviously, the additional polarization for the polaritons in comparison to the case of photons travelling in vacuum, results from the inclusion of a mechanical degree of freedom in the model. The three independent polarizations are consistent with the massive vector boson character of the polariton modes in the bulk as discussed in Sec. II B.

### D. Microscopic description of the polarization

In the previous sections we have used a phenomenological parameter $\alpha_B$ to characterize the response of the material to an electromagnetic perturbation. This phenomenological approach allows us to determine the possible polarizations of the modes and to classify the dispersion relation of each polarization. This macroscopic, phenomenological picture contains two unknowns: the bulk polarizability $\alpha_B$ and the refractive index $n = k/\omega$ (dispersion relation). We now need to find the relation between these quantities. In order to find the relation between bulk polarizability and the speed of propagation of the waves we need to consider the induced polarization at the microscopic level. The bulk polarizability is determined by the 'molecular' polarizability of the individual dipole oscillators considered in the model together with the interactions between the oscillators.

We start by considering the Lagrangian of an individual dipole oscillator:

$$L_{osc}(t,x,\dot{x}) = \frac{1}{2}m\dot{x}^2 - j_\alpha A^\alpha = \frac{1}{2}m\dot{x}^2 - qv_\alpha A^\alpha \tag{26}$$

Here we have used $j_\alpha = qv_\alpha$ for the microscopic current. In (26) we do not include any coupling with other mechanical modes such a low frequency lattice modes. We also do not include any terms related to radiative damping. The reason for this latter omission is that in a closed, linear system, dissipation of power can only result from coupling of the mechanical degree of freedom of the charged oscillators with other internal modes such as e.g. lattice modes. So if one chooses to ignore the coupling with lattice modes, then also no explicit damping terms need to be considered for closed systems where energy cannot leave. The equation of motion for the oscillator that can be derived from the Lagrangian reads:

$$m\ddot{x} = -q\left(\frac{\partial A_0}{\partial x}\right) + q\left(\frac{\partial A_x}{\partial t}\right) \tag{27}$$

Below we now work out this relation further, following the two-level course graining approach adopted. We distinguish short-range and long-range contributions to the potential $A_0$. The short range contribution includes the Coulombic interaction of the moving charge with is stationary counter charge located at the lattice point. This Coulombic interaction provides the restoring force for the harmonic oscillation of the moving charge:

$$\frac{\partial A_0^{short}}{\partial x} = -Kx \tag{28}$$

where $K$ is the force constant and $\omega_D^2 = K/m$ the square of the mechanical resonance frequency of an isolated oscillator.

If the oscillators are close together, the short range potential of one oscillator will also influence the potential at neighboring lattice point. This mutual interaction between neighboring oscillators can in first approximation be described by dipole-dipole interactions. The dipole-dipole interactions will in general lead to a shift of the resonance frequency. For our specific case of a cubic lattice, we note that in the case where all the dipoles have the same phase, unretarded dipole-dipole interactions cancel out [33]. In the two-level course graining approach we follow here, we assume that the wavelengths of the bulk polaritons are long compared to the lattice spacing $\Delta$, such that the phases of the dipoles on the local scale can be taken equal. Under this condition, a renormalization of the resonance frequency of the oscillators may be neglected for cubic lattices. Thus we approximate the resonance frequency of the coupled oscillators on the lattice ($\omega_Y$) by the resonance frequency $\omega_D$ for an isolated oscillator. We note that formally a renormalization of the resonance frequency, as required for e.g. non-cubic lattices, can be performed using Ewald's approach [34].

In addition to the short range contributions to the potential $A_0$, we also distinguish long range contributions associated with the wave character of the polaritons. Adopting the Lorentz gauge and assuming plane wave character for the polaritons $A \propto exp(i\omega t - i\vec{k}\cdot\vec{r})$ we get:

$$\frac{\partial A_0^{long}}{\partial x} = -i\frac{k_x k_x}{\omega} A_x - i\frac{k_x k_y}{\omega} A_y - i\frac{k_x k_z}{\omega} A_z \tag{29}$$

The combined equation of motion for the oscillators now reads:

$$\frac{m}{q}\ddot{x} = -\frac{K}{q}x + i\frac{k_x k_x}{\omega} A_x + i\frac{k_x k_y}{\omega} A_y + i\frac{k_x k_z}{\omega} A_z - i\omega A_x \tag{30}$$

We consider now excitation of a 'heavy' mode for which $A_y = A_z = 0$. We assume harmonic periodic oscillations $x(t) = x0\, e^{i\omega t}$, corresponding to equilibrium between mechanical and electromagnetic degrees of freedom. This gives

$$x = -\frac{iq}{\omega m} \frac{\omega^2 - k_x^2}{\left(\omega_D^2 - \omega^2\right)} A_x \tag{31}$$

$$j_x = q\dot{x} = \frac{q^2}{m} \frac{\omega^2 - k_x^2}{\left(\omega_D^2 - \omega^2\right)} A_x \tag{32}$$

where $\omega_D$ is now the resonance frequency of the isolated dipole oscillator. From (32) we can now extract an expression for the microscopic polarizability $\alpha_M$:

$$\alpha_M(\omega, \vec{k}) = \frac{q^2}{m} \frac{\omega^2 - k_x^2}{\left(\omega_D^2 - \omega^2\right)} \tag{33}$$

Importantly, we note that by describing the electromagnetic degrees of freedom by vector potentials, we have already incorporated magnetic dipole terms in the interaction between the oscillators and electromagnetic fields. Hence spatial dispersion, *i.e.* a dependence of the polarizability on the wave vector *k*, is automatically included [35].

### E. A scattering approach

Having now established the precise functional dependence of the microscopic response, we now need to relate the polarization of the individual sites to the macroscopic polarization. This relation can be found by realizing that each individual oscillator responds to the combined fields of the incoming primary electromagnetic wave and the secondary radiation waves from all other dipole oscillators. Self-consistency can then be demanded to find the combined total field. In this section we give a short approximate, intuitive solution to the problem of finding the effective field based on insights from Rayleigh [36] and Sellmeier [37,38]. Rayleigh and Sellmeier recognized that the index of refraction of a medium containing scattering centers with density $N$ is related to the amplitude for forward scattering $f(0)$ of the individual scatterers. This relation expressed in gaussian units, is well known:

$$n = 1 + 2\pi N \frac{f(0)}{k^2} \tag{34}$$

where $k$ is the wavevector in the incident direction. Importantly, the forward scattering amplitude is related to the polarizability of the scatterers. This relation can be derived in the scalar wave approximation [39]. We consider a slab of material, extending from 0 to $y$ in $y$-direction and to $\pm\infty$ in the $x$- and $z$-directions and a beam incident along the $y$-direction. The transmitted wave at position $y$ can be expressed as the sum of the primary beam and scattered waves

$$u = u_0 \left\{ e^{-ik_{vac}y} + \alpha_M \sum \frac{1}{4\pi r} e^{-ik_{vac}r} \right\} = u_0 e^{-ik_{vac}y} \left\{ 1 + \frac{\alpha_M}{4\pi\Delta^3} \int_0^y \int_{-\infty}^{\infty} \int_{-\infty}^{\infty} dy\,dx\,dz \frac{1}{r} e^{ik_{vac}\frac{x^2+z^2}{2y}} \right\} \tag{35}$$

where the sum runs over all scattering centers in the slab. The amplitude of the wave scattered by an individual dipole is taken as $\alpha_M e^{-ik_{vac}r}/r$ where $r$ is approximated as $r \cong y + (x^2 + z^2)/2y$

$$u = u_0 \exp\left(-ik_{vac}\left[1 + \frac{\alpha_M}{2\Delta^3}\frac{1}{k_{vac}^2}\right]y\right) \tag{36}$$

This relation shows that a phase delay accumulates for the wave travelling through the material. The phase delay can be incorporated into a refractive index by expressing the wavevector of the light propagating in the slab as $k_{med} = k_{vac} n$. The resulting approximate expression for the refractive index is then

$$n = 1 + \frac{1}{2\Delta^3}\frac{\alpha_M}{\omega^2} \tag{37}$$

We note that this relation between molecular polarizability and the refractive index of the medium differs from the Lorenz-Lorentz expression. As has been discussed [39], the Lorenz-Lorentz expression is not rigorous. The two relations converge for low oscillator densities.

**F. Microscopic derivation of the relation between refractive index and molecular polarizability.**

A rigorous derivation of the relation between refractive index and molecular polarizability is possible by summing up all the scattered components of the field. We focus on the $x$-component of the vector potential which we now simply abbreviate as $A$. We suppress the time factor $e^{-i\omega t}$ throughout this section. The total amplitude of the potential at point $x$ can be expressed as the sum of an incoming field ($A_{\text{inc}}$) and waves scattered off the dipoles ($A_{\text{dip}}$):

$$A_{\text{eff}}(x) = A_{\text{inc}}(x) + \sum_j A_{\text{dip},j}(x) \tag{38}$$

Note that in (38) the effective field at site $i$ also includes a contribution from the dipole field of the oscillator at site $i$. As discussed in Sec. II D, the short-range components of the effective field around a lattice site provide the restoring force for the harmonic oscillation of the electron. We argue therefore that the on-site dipole fields should be included in the effective field, because they provide the necessary short range components.

The amplitude of the field scattered off dipole $i$ located at $x_i$ and evaluated at point $x$ can be expressed as:

$$A_{\text{dip},i}(x) = \alpha_M G(x, x_i) A_{\text{eff}}(x_i) \tag{39}$$

where we have introduced the spatial Green function:

$$G(x, x_i) = -\frac{1}{4\pi} \frac{e^{-ik_{\text{vac}}|x-x_i|}}{|x-x_i|} \tag{40}$$

satisfying $\Box G = -\delta$ and with $k_{vac} = \omega$ the wavevector obeying the vacuum dispersion relation.

Combining (38) and (39) we get:

$$A_{eff}(x_i) = A_{inc}(x_i) + \sum_j \alpha_M G(x_i, x_j) A_{eff}(x_j) \tag{41}$$

For the incoming field coming from a source far away from the molecular material we assume vacuum dispersion $\propto e^{-ik_{vac}x}$. For the effective field inside the medium we take the Ewald-Oseen Ansatz: $\propto e^{-ik_{eff}x}$. The effective field can now be expressed as:

$$A_{eff}(x_i) = \frac{A_{inc} e^{-i(k_{vac}-k_{eff})x_i}}{\left[1 - \sum_j \alpha_M G(x_i, x_j) e^{-ik_{eff}(x_j-x_i)}\right]} \stackrel{(x_i=0)}{=} \frac{A_{inc}}{[1 - \alpha_M S_0]} \tag{42}$$

where the origin has been chosen to coincide with the lattice point $x_i$ and where we have introduced the sum:

$$S_0 = \sum_j G(0, x_j) e^{-ik_{eff}(x_j)} \tag{43}$$

A simple approach to calculating this sum is to approximate it by an integral, using the well known plane wave expansion of the Yukawa potential:

$$-\frac{e^{-ikr}}{4\pi r} = \int \frac{d^3k'}{(2\pi)^{3/2}} \frac{e^{i\vec{k}'\cdot\vec{r}}}{k^2 - k'^2} \tag{44}$$

Strictly, this expansion is only convergent if $k'$ has an imaginary component. Convergence can be enforced by introducing and imaginary component $i\varepsilon$ and later taking the limit $\varepsilon \to 0$. Alternatively, a somewhat more elaborate derivation is possible making use the Weyl expansion of the spherical wave [30].

$$S_0 = \frac{1}{\Delta^3} \frac{1}{k_{vac}^2 - k_{eff}^2} \tag{45}$$

If we translate the point of observation over one lattice spacing Δ, then the effective field can only differ by a phase factor:

$$A_{eff}(x+\Delta) = A_{eff} e^{-ik_{eff}\Delta} \tag{46}$$

This expression can be inserted in (41) to give:

$$A_{inc} + A_{eff}\alpha_M S_0 = A_{inc} e^{-ik_{vac}\Delta} + A_{eff} e^{-ik_{eff}\Delta} \alpha_M S_0 \tag{47}$$

From (47) it follows that for lattice spacings smaller compared to the wavelength:

$$\left[\frac{\alpha_M S_0}{1-\alpha_M S_0}\right] = \frac{-\left(1-e^{-ik_{vac}\Delta}\right)}{\left(1-e^{-ik_{eff}\Delta}\right)} \cong -\frac{k_{vac}}{k_{eff}} \tag{48}$$

By inserting the full expression for the sum $S_0$ (45) into (48) it follows [30]:

$$\frac{k}{\omega} = 1 + \frac{1}{\Delta^3} \frac{\alpha_M}{\omega^2} \tag{49}$$

Note that the expression (49) is similar to (37), differing by only a factor of two in the second term. This difference may be due to the fact that in the approximate treatment of Sec. II E only forward scattering has been considered and contributions from backward scattering have been ignored.

## G. The dispersion relation for polaritons in a uniaxially polarizable medium

### 1. 'Heavy' $k_x=0$ mode

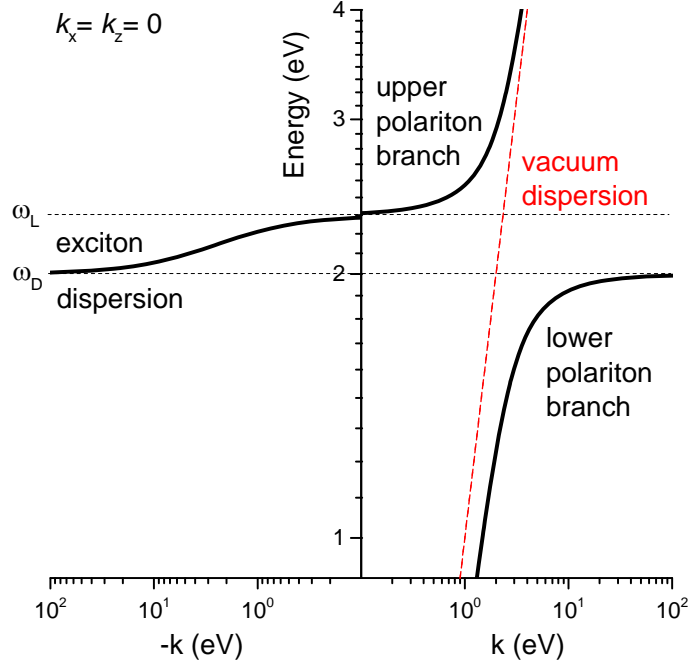

**FIGURE 2.** Dispersion relation for 'heavy' polaritons with $k_x = k_z = 0$ and model parameters $\Delta = 0.005$ eV$^{-1}$ ($\cong 1$ nm) and $\omega_D = 2$ eV. The red dashed line gives the dispersion for photons in vacuum as reference.

In the previous section we established the relation between refractive index and the molecular polarizability. This allows us know to investigate the dispersion relations for the polaritons in detail. Referring back to Table I, we note that the 'light' polariton modes follow a trivial dispersion relation ($\omega = k$). The dispersion relations for the 'heavy' polariton modes are far more complex. We start with the case where the spatial part of the wavevector is perpendicular to the direction of the charge oscillators ($k_1 = 0$). In this case there is only one 'heavy' polariton. The dispersion relation for this polariton can be obtained by inserting the expression for the molecular polarizability (33) into the expression (49) for the refractive index. In Fig. 2 we have plotted this dispersion relation for the case $\omega_D$ = 2 eV and a lattice constant $\Delta$ = 0.005 eV$^{-1}$, corresponding to 1 nm in SI units.

$$\frac{k}{\omega} = 1 + \frac{1}{\Delta^3} \frac{\alpha_M(\omega, k_x = 0)}{\omega^2} = 1 + \frac{q^2}{\Delta^3 m} \frac{1}{(\omega_D^2 - \omega^2)} = 1 + \frac{\omega_P^2}{(\omega_D^2 - \omega^2)} \quad (50)$$

Looking at Fig. 2, we first discuss the low frequency range $\omega < \omega_D$. In the limit of very low frequencies the dispersion curve runs parallel to the dispersion relation for photon in vacuum. In the graph the curve for the polaritons lies the photon line because, at a particular frequency the magnitude of the wavevector for the polariton equals the magnitude of the vacuum wavevector multiplied by the refractive index. At low frequency the refractive index does not converge to zero but approached a finite value. Extrapolating to zero frequency, we obtain a static dielectric constant

$$\varepsilon_{xx} = n^2(\omega = 0) = \left(1 + \frac{\omega_P^2}{\omega_D^2}\right)^2 \quad (51)$$

In the low frequency region or so-called lower polariton branch, the resonance frequency of the isolated oscillators $\omega_D$ serves as an asymptotic limit for the frequency when the wavevector is raised.

For frequencies in the interval between $\omega_D$ and $\omega_L$ the wave vector is negative. We interpret the negative sign for the wave vector in terms of backscattering of the incoming wave. An expression for $\omega_L$ can derived from the dispersion relation (49):

$$\omega_L = \sqrt{\omega_D^2 + \omega_P^2} \quad (52)$$

For frequencies above $\omega_L$ the dispersion curve converges to the photon dispersion line. At high frequency the refractive index approaches unity.

### 2. 'Heavy' $k_x \neq 0$ modes

The dispersion relation for 'heavy' polaritons whose wave vector has a non-zero component in the direction of the charge oscillators is more complex. For this case Eq. 25 reduces to

$$\begin{bmatrix} kk - n^2\alpha_B & \alpha_B n & 0 & 0 \\ -n\alpha_B & kk + \alpha_B & 0 & 0 \\ 0 & 0 & kk & 0 \\ 0 & 0 & 0 & kk \end{bmatrix} \begin{bmatrix} nA_1 \\ A_1 \\ 0 \\ 0 \end{bmatrix} = 0 \tag{53}$$

The first two rows in this equation are in fact identical. This identity roots in charge conservation on the long scale which demands $k_1 J_1 = \omega J_0$. Based on this equivalence, the microscopic contributions of the dipole oscillators to the vector potential for the case where the wave vector has a component in the $x$-direction can be expressed as:

$$A_{eff,1}(x_i) = A_{inc,1}(x_i) - \sum_j \alpha_M G(x_i, x_j) A_{eff,1}(x_j) \tag{54a}$$

Following directly from $\partial^\alpha A_\alpha = 0$ we have:

$$\frac{A_{eff,0}}{k_x^{eff}}(x_i) = \frac{A_{inc,0}}{k_x^{vac}}(x_i) - \sum_j \alpha_M G(x_i, x_j) \frac{A_{eff,0}}{k_x^{eff}}(x_j) \tag{54b}$$

which corresponds to (41).

Combining (49) and (33) yields:

$$\frac{k}{\omega} = 1 + N\frac{\alpha_M}{\omega^2} = 1 + \frac{q^2}{m\Delta^3} \frac{\omega^2 - k^2 d^2}{\omega^2(\omega_D^2 - \omega^2)} = 1 + \omega_P^2 \frac{\omega^2 - k^2 d^2}{\omega^2(\omega_D^2 - \omega^2)} \tag{55}$$

where $k$ is the magnitude of the spatial wavevector $\vec{k}$ and $d$ is the sine of the angle between $\vec{k}$ and the $y$ axis.

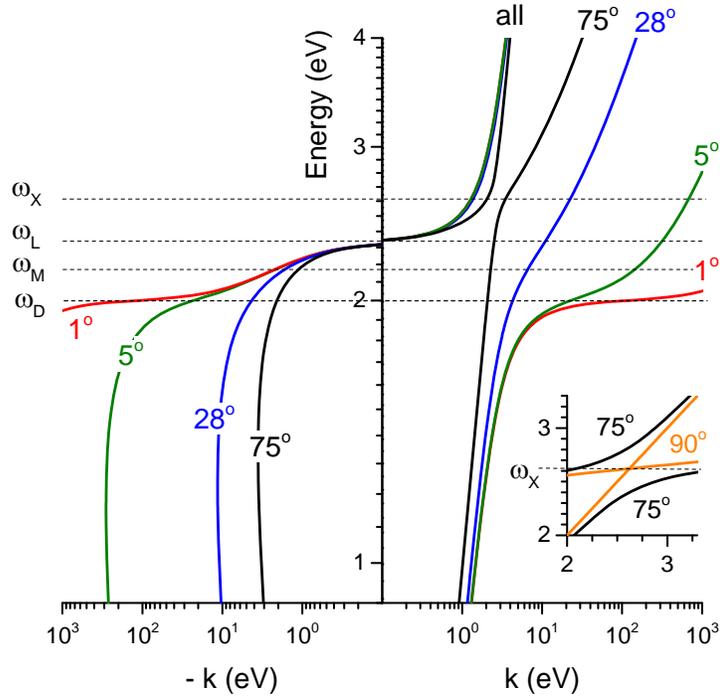

**FIGURE 3**. Dispersion relation for 'heavy' polaritons for different directions of the wavevector with $k_x \neq 0$, $k_z = 0$. Numbers in degrees indicate the angle $\theta_k$ between the $\vec{k}$ vector and the $y$-direction. The inset illustrates that for polaritons with $\vec{k}$ vector parallel to the orientation direction of the oscillators ($\theta_k = 90°$), the splitting between the upper and lower polariton branches vanishes and the exciton and photon dispersion lines cross at $\omega_X = 2.621$ eV.

In Fig. 3 we illustrate the dispersion relation (55) for different directions of the wave-vector $\vec{k}$. For small angles $\theta_k = 1°$ between the direction of the $k$ vector and the $y$ axis, the dispersion curve closely resembles the case of heavy mode with $k_x = 0$ illustrated in Fig. 2.

For large angle $\theta_k$, we find that for each energy $\omega$, two possible polariton modes with different $k$ value exist. This is consistent with the earlier analysis of the parameter space, see Table I. For $\omega \gg \omega_D$ these two modes can be interpreted as a photon like polariton with small k on the upper polariton branch and an exciton like polariton with large $k$ on lower polariton branch. We note that these exciton-like polaritons follow approximately a quadratic dependence of the energy on wave vector for small angles. This quadratic dependence is consistent with the absence of linear dispersion terms in materials with inversion symmetry.

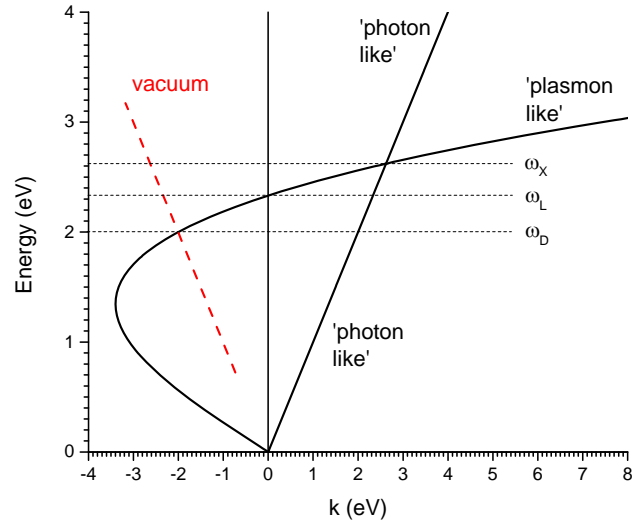

**FIGURE 4**. Dispersion relation for polaritons with $\vec{k}$ vector parallel to the orientation direction of the oscillators ($\theta_k = 90°$, $k_y = k_z = 0$) and polarization in the $xy$ plane. Model parameters: lattice constant $\Delta = 0.005$ eV$^{-1}$ and dipole oscillator resonance frequency $\omega_D = 2.0$ eV. The 'plasmon like' branch crosses the dispersion line for photons in vacuum at $\omega = \omega_D$ and at $\omega = \omega_X = 2.621$ eV.

For large angles $\theta_k$, e.g 75° the upper and lower polariton branches come close together. The minimum distance between the upper and lower polariton branch is a measure for the strength of the interaction between photon and exciton. For an angle $\theta_k = 90°$ the curves actually cross at $\omega_X = \sqrt{\omega_D^2 + 2\omega_P^2}$ (see insert Fig. 3). The crossing indicates that the exciton-photon coupling vanishes. Thus for wave-vectors parallel to the $x$-direction, pure photon and pure exciton modes coexist without any hybridization to polaritons. In Fig. 4 we illustrate the dispersion relations for this case in more detail. As expected, the photon obeys the vacuum dispersion relation as indicated by the straight line. The second type of excitation in this case, the exciton, consists of a mechanical-electrical polarization in the *x*-direction propagating in the same direction. Therefore the exciton can be considered as a longitudinal wave. The dependence of the frequency of the longitudinal excitons on the wavevector shows a square root behavior, similar to the case of surface polaritons on metals.

Looking again at the general dispersion curves in Fig. 3, we note that for $\omega \ll \omega_D$ also two solutions exist; one 'heavy' polariton mode with positive *k* value and another 'heavy' polariton with negative *k*. The negative *k* solution we interpret, in analogy to the negative *k* solutions in Fig. 2, as a coherently backscattered wave.

In summary, the two-scale modeling approach followed here leads to dispersion relations that predict the presence of three independent polarizations for excitations in an anisotropic medium propagating in a particular direction. The existence of three mechano-electro-magnetic modes inside anisotropic media has been presented earlier in literature [35] and has been studied in considerable detail for chiral or gyrotropic media [40]. Also for cholesteric liquid crystals the existence of more than two electromagnetic modes has been discussed [41]. The model presented here indicates that the presence of more than two electromagnetic modes might be quite general and can also occur for frequencies below the optical resonance frequency of the dielectric materials.

**H. Boundary conditions for the vector potential at the interface between vacuum and bulk polarizable matter**

In general, coupled mechanical-electromagnetic excitations can have up to *three* independent components, see Sec. II C. Importantly, in analyzing reflection and refraction, it should be noted that the presence of polarizable matter breaks the gauge symmetry also in the vacuum part of the problem. One has to insist on uniform gauge conditions throughout the entire space considered in the problem under study. In practice this means the popular Coulomb gauge for light propagating on the vacuum side of the problem cannot be used. In this section we derive boundary conditions for the vector potential at the interface between vacuum and matter, taking into account gauge.

We consider a planar interface located at $y = 0$ between vacuum and a cubic lattice of dipole oscillators as discussed in Sec. IIA. The dipoles are oriented in the $x$-direction.

Formally, the variation in vector potential $A$ across the interface $y = 0$ can be described using the singular step function $S(y)$ [42-44]:

$$A^\beta = A^\beta_{|a} + \left( A^\beta_{|b} - A^\beta_{|a} \right) S(y) \tag{56}$$

Here the subscripts $|a$ and $|b$ indicate, resp., the vector potential in vacuum and in the material.

Analogously the current density can be expressed as:

$$J^\beta = J^\beta_{|a} + \left( J^\beta_{|b} - J^\beta_{|a} \right) S(y) + J^\beta_{|s} S'(y) = J^\beta_{|b} S(y) + J^\beta_{|s} S'(y) \tag{57}$$

Here we have used the fact that the current density in vacuum must be zero. $S'$ stands for the first derivative of the step function in the $y$ direction [45], which is equivalent to a delta function. $J^\alpha_{|s}$ indicates a two dimensional current density *at* the interface that may be needed to ensure charge conservation. In our case, because the dipoles in the material are oriented in a direction parallel to the surface, the specific surface current density $J^\alpha_{|s}$ equals zero.

To derive the boundary conditions for the vector potential at the interface, we re-evaluate the Euler-Lagrange equation (6) by substituting the expression (56) for the vector potential $A$.

$$\partial_\beta F^{\beta\alpha} = \partial_\beta \left( \partial^\alpha A^\beta - \partial^\beta A^\alpha \right) = J^\alpha \tag{58}$$

We note that the first term in Eq. 58 $A^{\alpha}_{|a}$, describing the vector potential in vacuum, does not make a contribution because the corresponding current density in vacuum $J^{\alpha}_{|a}$ exactly equals zero. We focus therefore on the second term on the RHS of (56) $(A^{\alpha}_{|a} - A^{\alpha}_{|b})S$, and rewrite (58) introducing the shorthand $\bar{A}^{\alpha} = A^{\alpha}_{|b} - A^{\alpha}_{|a}$:

$$\partial_{\beta}\partial^{\alpha} S\bar{A}^{\beta} - \Box S\bar{A}^{\alpha} = J^{\alpha}_{|b} S(y) \tag{59}$$

The first term on the LHS of (59) can be worked out further making use of the Lorentz gauge condition $\partial_{\beta}A^{\beta} = 0$:

$$\partial_{\beta}\left(\partial^{\alpha}\left(S\bar{A}^{\beta}\right)\right) = S\overline{k^{\alpha}\left(\partial_{\beta}A^{\beta}\right)} + S'_{y}\left(\partial_{\beta}\bar{A}^{\beta}\right) + \overline{k^{\alpha}A^{y}}S'_{y} + \bar{A}^{y}S''_{y} \tag{60}$$

Where we have used $\partial_{\alpha}S = \partial_{y}S = S'_{y} = -S'^{y}$, $\partial^{y}S = S'^{y}$ and $\partial_{y}\partial^{y}S = S''^{y}_{y}$

Also the second term on the LHS of (59) can be worked out:

$$-\Box S\bar{A}^{\alpha} = -S\Box\bar{A}^{\alpha} - 2S'_{y}\overline{k^{y}A^{\alpha}} - \partial_{\beta}\bar{A}^{\alpha}S''^{y}_{y} \tag{61}$$

Combined we get:

$$-S\Box\bar{A}^{\alpha} + S'_{y}\left(\overline{k^{\alpha}A^{y}} - 2\overline{k^{y}A^{\alpha}}\right) + S''^{y}_{y}\left(\bar{A}^{y} - \bar{A}^{\alpha}\right) = SJ^{\alpha}_{|b} \tag{62a}$$

The relation (62a) can be broken down into three separate parts pertaining to $S$, $S$' and $S$''

$$\begin{cases} -S\Box\bar{A}^{\alpha} = SJ^{\alpha} \\ S'_{y}\left(\overline{k^{\alpha}A^{y}} - 2\overline{k^{y}A^{\alpha}}\right) = 0 \\ S''^{y}_{y}\left(\bar{A}^{y} - \bar{A}^{\alpha}\right) = 0 \end{cases} \tag{62b}$$

The first of these three equations covers the behavior of both the bulk, $-\Box A^{\alpha} = J^{\alpha}$, and the vacuum, $-\Box A^{\alpha} = 0$. The second and third equations in (62b) specify boundary conditions for the vector potential at the interface:

$$\begin{cases} \left(k_{|b}^{\alpha} A_{|b}^{y} - 2k_{|b}^{y} A_{|b}^{\alpha}\right) - \left(k_{|a}^{\alpha} A_{|a}^{y} - 2k_{|a}^{y} A_{|a}^{\alpha}\right) = 0 \\ A_{|b}^{y} - A_{|a}^{y} - A_{|b}^{\alpha} + A_{|a}^{\alpha} = 0 \end{cases} \tag{63a}$$

The relations (63a) should hold for each of the four components of the four vector $A^{\alpha}$:

$$\alpha = 0 \quad t \quad \begin{cases} \left(\left(\omega A_{|b}^{y} - 2k_{|b}^{y}\Phi_{|b}\right) - \left(\omega A_{|a}^{y} - 2k_{|a}^{y}\Phi_{|a}\right)\right) = 0 \\ \left(A_{|b}^{y} - A_{|a}^{y} - \Phi_{|b} + \Phi_{|a}\right) = 0 \end{cases}$$

$$\alpha = 1 \quad x \quad \begin{cases} \left(\left(k_{|b}^{x} A_{|b}^{y} - 2k_{|b}^{y} A_{|b}^{x}\right) - \left(k_{|a}^{x} A_{|a}^{y} - 2k_{|a}^{y} A_{|a}^{x}\right)\right) = 0 \\ \left(A_{|b}^{y} - A_{|a}^{y} - A_{|b}^{x} + A_{|a}^{x}\right) = 0 \end{cases}$$

$$\alpha = 2 \quad y \quad \begin{cases} \left(-k_{|b}^{y} A_{|b}^{y} + k_{|a}^{y} A_{|a}^{y}\right) = 0 \\ 0 = 0 \end{cases} \tag{63b}$$

$$\alpha = 3 \quad z \quad \begin{cases} \left(\left(k_{|b}^{z} A_{|b}^{y} - 2k_{|b}^{y} A_{|b}^{z}\right) - \left(k_{|a}^{z} A_{|a}^{y} - 2k_{|a}^{y} A_{|a}^{z}\right)\right) = 0 \\ \left(A_{|b}^{y} - A_{|a}^{y} - A_{|b}^{z} + A_{|a}^{z}\right) = 0 \end{cases}$$

Where $\Phi$ indicates the time-like component of the vector potential that can be interpreted as the electrostatic potential. Importantly, we note that (63) cannot be derived by directly combining (7) and (56) because then the last two terms contained in (61) involving derivatives of $S$ would be left out. This illustrates the importance of proper handling of the gauge conditions. For the case of a perfect electrically conducting medium, it can be shown that the relation (63b) reduces to the well-known Drude expression for the normal reflection of light from a metal [30].

We note that in our derivation of the boundary conditions for the vector potential, one is in principle free to choose any gauge condition. However, when deriving (63) the Lorentz gauge was adopted, and so in applications of (63) this gauge condition should be adhered to.

## I. Reflectance

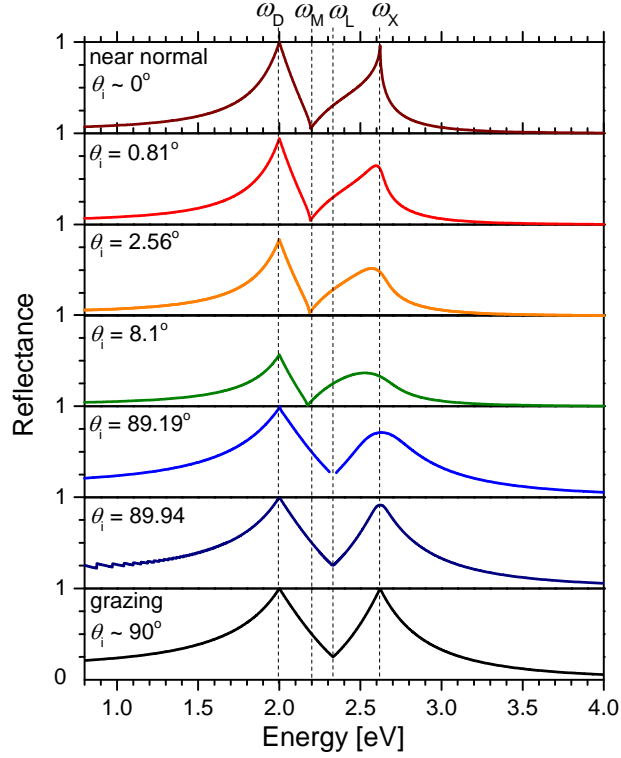

**FIGURE 5**. Reflectance versus photon energy for different angles of incidence $\theta_i$ onto the surface of a uniaxially polarizable medium with $\Delta = 0.005$ eV$^{-1}$ and $\omega_D = 2$ eV. The incoming light is polarized parallel to the plane of incidence (*xy*). The dipole oscillators are oriented in the *x*- direction, perpendicular to the surface normal (-*y*).

In this section we consider in detail the reflection of light from a surface of a uniaxially polarizable medium. The vacuum-matter interface is described by $y = 0$ and the oscillators are oriented in the *x*-direction, see Fig. 1. The reflection and refraction of light incident in the *xy* plane will be studied.

In the geometry chosen, the case of refraction of incoming light with polarization in the *z*- direction ($A_x = A_y = 0$) is trivial. Because the polarization of the light is always perpendicular to the dipole oscillators, there is no light-matter interaction. Reflection will thus be zero and the angle of refraction equals the angle of incidence.

Next we study the reflection and refraction of light with polarization in the *xy* plane of incidence (*p*-polarization). The presence of the polarizable medium breaks the gauge symmetry of the photons. The key to finding a solution for reflection and refraction is the notion that one needs to adopt a uniform gauge for both the vacuum and the matter half spheres. The gauge chosen needs to cover the broken gauge symmetry in matter and should therefore itself be broken. We first consider the polariton modes inside the material. Here clearly the interesting case involves the 'heavy' polariton modes. Electromagnetic waves inside the medium with $A_y$ polarization obey vacuum dispersion, hence does not interact with the lattice. Therefore, *y*-polarized waves are less interesting. The interesting 'heavy' polariton modes can be selected by choosing the Lorentz gauge, with the condition that the $A_y$ component of the modes is zero. As explained in Sec. II.B, because the $A_y$ modes travel with vacuum dispersion, they can always be 'gauged out' from the problem. To ensure uniformity of gauge throughout the problem, one is forced to adopt a rather unconventional gauge condition for the electromagnetic radiation on the vacuum side of the problem which is based on the Lorentz gauge with the additional condition that the $A_y$ component is set to zero. Considering that the light in vacuum travelling in a particular direction must have two independent polarizations, it follows that the complementary polarization must be in the *z*-direction. For this second type of polarization, the reflection and refraction is trivial, see above.

Using now the gauge requirement that the *y*-component of the vector potential must be zero on both sides of the interface, the boundary conditions (63b) for the vector potential at the interface simplify to:

$$t \begin{cases} k_{|1}^y \Phi_{|1} + k_{|2}^y \Phi_{|2} = k_I^y \Phi_I + k_R^y \Phi_R \\ \Phi_{|1} + \Phi_{|2} = \Phi_I + \Phi_R \end{cases}$$

$$x \begin{cases} k_{|1}^y A_{|1}^x + k_{|2}^y A_{|2}^x = k_I^y A_I^x + k_R^y A_R^x \\ A_{|1}^x + A_{|2}^x = A_I^x + A_R^x \end{cases} \quad (64)$$

$$y \begin{cases} 0 = 0 \\ 0 = 0 \end{cases}$$

$$z \begin{cases} 0 = 0 \\ 0 = 0 \end{cases}$$

On the vacuum side there will be an incoming (I) and reflected (R) wave. On the material side, as discussed in Sec. II C, are two 'heavy' polariton modes, labeled |1 and |2 that can be excited by the incoming light. The magnitudes of the potential for each wave follows from the Lorentz gauge condition:

$$\begin{aligned} \omega \Phi_{|1} &= k_{|1}^x A_{|1}^x \\ \omega \Phi_{|2} &= k_{|2}^x A_{|2}^x \\ \omega \Phi_I &= k_I^x A_I^x \\ \omega \Phi_R &= k_R^x A_R^x \end{aligned} \quad (65)$$

At a given frequency, the *y*-component of the wave vector of the incident ray, $k_I^y$ can be considered as a free parameter containing the information on the angle of incidence. The magnitude of the *x*-component follows from the vacuum dispersion relation:

$$\left| k_I^x \right| = \sqrt{\omega^2 - \left(k_I^y\right)^2} \quad (66)$$

For the reflected ray, the requirement of phase matching dictates $k_R^y = -k_I^y$. The component $k_R^x$ follows analogy with (66).

Considering now the medium, we note that because of the uniform orientation of the dipole oscillators in the *x*-direction, there can be no contribution for the mechanical degree of freedom to the momentum of any wave in the *y*-direction. Hence, we argue that the *y*-components of the momenta of the two heavy polaritons must be equal $k_{|1}^y = k_{|2}^y$. The *x*-components of the momenta can differ but should be related to the corresponding *y*-component via the dispersion relation.

$$t \begin{cases} k_I^y k_{|1}^x A_{|1}^x + k_I^y k_{|2}^x A_{|2}^x = k_I^y k_I^x \left( A_I^x - A_R^x \right) \\ k_{|1}^x A_{|1}^x + k_{|2}^x A_{|2}^x = k_I^x \left( A_I^x + A_R^x \right) \end{cases}$$

$$x \begin{cases} k_{|1}^y A_{|1}^x + k_{|1}^y A_{|2}^x = k_I^y \left( A_I^x - A_R^x \right) \\ A_{|1}^x + A_{|2}^x = A_I^x + A_R^x \end{cases} \tag{67}$$

Obviously, relation (67) on its own, so without taking into account the dispersion relation (49), admits a trivial solution with $|k_{|1}| = |k_{|2}| = |k_I| = \omega$ and $A_R^x = 0$. Relation (67) combined with (49) also admits nontrivial solutions. In order to describe these solutions we first introduce two auxiliary functions, *f* and *g*:

$$g = \frac{k_I^y - k_{|1}^y}{k_I^y + k_{|1}^y} = \frac{K-1}{K+1}$$

$$f = \frac{k_I^x - k_{|1}^x}{k_{|2}^x - k_I^x} \tag{68}$$

where $K = k_I^y / k_{|1}^y$. The four relations contained in (67) can now be rewritten as [30]:

$$\begin{cases} A_R^x = g A_I^x & (a) \\ A_{|1}^x = \frac{1+f}{1+g} A_I^x & (b) \\ A_{|2}^x = \frac{f + f^2}{1+g} A_I^x & (c) \\ \left( k_{|2}^x - k_{|1}^x \right)^2 (K+1)^2 = \left( k_{|2}^x - k_I^x \right)^2 (2K)^2 & (d) \end{cases} \tag{69}$$

The three relations 69a,b and c specify the field amplitudes. The fourth expression, Eq. 69d relates the *y*-components of the wavevector of the refracted rays ($k_{|1}^y$) and the *y*-component of the incoming radiation ($k_I^y$). The latter quantity is determined by the experiment. Also $k_I^x(\omega, k_I^y)$ is fixed and given by (66). Thus the unknown quantity that needs to be extracted from (69d) in order to calculate the reflected and refracted rays is $k_{|1}^y$. We note that the *x*-components of the wave-vectors of the refracted rays $k_{|2}{}^x(\omega, k_{|1}{}^y)$ and $k_{|2}{}^x(\omega, k_{|1}{}^y)$ in (69d) are functions of $k_{|1}{}^y$ due to (49). Thus, given a frequency and direction of the incoming beam specified by e.g. the set of parameters $(\omega, k_I{}^y)$, the wave-vectors and amplitude for the two refracted rays can be calculated by first determining $k_{|1}{}^y$.

The strategy that we follow to solve (69d) is as follows. We first note that in the limits of grazing and near-normal incidence, (69d) can be simplified. Approximate solutions [30] for our standard oscillator lattice ($\Delta = 0.005$ eV$^{-1}$ and $\omega_D = 2.0$ eV) are shown in Fig. 5. We then use numerical root searching to find solutions for both small and large angles of incidence. Additional requirements imposed on the solutions are reflectivities smaller or equal than unity, a vanishing reflectance in the limit $\omega \to \infty$ and finite field amplitudes.

The predicted spectra shown in Fig. 5 feature two maxima of reflectance. One of the maxima is located near dipole oscillator frequency $\omega_D$ and the other near the crossing frequency $\omega_X$ (see Fig. 3), defined by:

$$\omega_X^2 = \omega_D^2 + 2\omega_P^2 \tag{70}$$

In the limits of gazing and normal incidence, the positions of the maxima coincide exactly with $\omega_D$ and $\omega_X$. The two maxima are separated by a minimum in reflectance. The position of the minimum on the frequency axis depends on the angle of incidence. For grazing incidence the minimum occurs at:

$$\omega_L^2 = \omega_D^2 + \omega_P^2 \tag{71}$$

For normal incidence the minimum occurs at

$$\omega_M^2 = \omega_D^2 + \frac{\omega_P^2}{2} \tag{72}$$

## J. Refraction

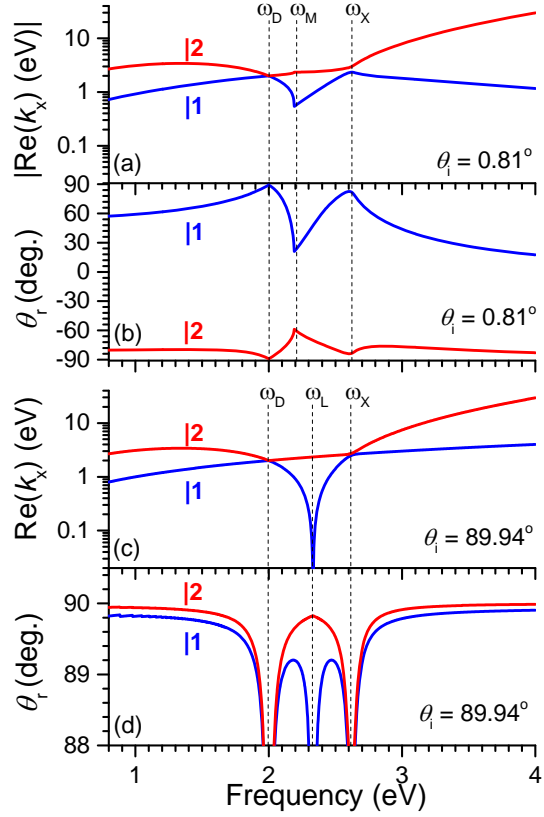

**FIGURE 6**. Waves refracted at the *xz* surface of the cubic lattice of dipole oscillators oriented in the *x*-direction for light incident in the *xy* plane with in-plane polarization. $\Delta = 0.005$ eV$^{-1}$ and $\omega_D = 2.0$ eV. (a) Absolute value of the real part of the *x*-component of the wavevector of the two heavy polaritons |1 and |2 excited by light incident with angle of incidence $\theta_i = 0.81°$. (b) Corresponding angle of refraction $\theta_r$ for $\theta_i = 0.81°$. (c,d) Corresponding plots for grazing incidence $\theta_i = 89.94°$.

Part of the light incident in the *xy* plane (on the *xz* interface) with in-plane polarization is refracted by the cubic lattice of dipole oscillators. Fig. 6a shows the *x*-components of the two 'heavy' polaritons labeled |1 and |2 that are excited by the incident light. The angles of refraction

for the two heavy polariton modes are shown in Fig. 6b. Interestingly, the angles of refraction for the two heavy polaritons modes are *different*. We recall that apart from the two heavy polaritons illustrated in Fig.6ab, also an electromagnetic wave polarized in the *z*-direction exists. For the latter *z*-wave, the angle of refraction equals the angle of incidence. This implies that an incoming unpolarized light beam is predicted to be refracted into three separate beams, each with a different polarization and each travelling in a different direction. This phenomenon might be described as '*trirefringence*'.

For anisotropic materials, the occurrence of *birefringence*, where an incoming beam is refracted into just *two* separate rays, is well established. Naturally, an important question with regards to the calculations presented here, is what the nature of the predicted additional third wave could be. We note that the 'heavy' polariton |2 in Fig. 6b propagates in a direction almost parallel to the matter/vacuum interface. In this respect, the polariton labeled |2 resembles a surface wave. Furthermore, in any dielectric, charge in the bulk of the material is accompanied by an associated 'image' charge at the surface. Therefore it may be expected that any electronic excitation travelling through the bulk of a dielectric should be accompanied by wave of charge at the surface [46]. The additional polariton wave that occurs in our model may therefore be interpreted as such surface wave accompanying an excitation in the bulk. The existence of surface waves at interfaces between non-metallic media is well supported both theoretically and experimentally. At the surfaces of chiral materials [47] and at the edges of photonic crystals, surface waves are known [48,49]. At the interface between anisotropic dielectric materials, so-called Dyakonov waves occur [50-52]. For structured anisotropic material interfaces Dyakonov-Tamm waves exist at the interface of two dielectric materials [53]. Experimental studies on the propagation of light in photonic waveguides have indeed provided evidence for the existence of trirefringence [54]. In addition, multirefringence is known to occur in nonlinear optics [55-57].

Fig. 6c,d illustrates the refracted wave for light incoming under grazing incidence. Also here we find two heavy polaritons travelling in different directions. In this case, the second polariton also mainly travels in a direction parallel to the surface. Exceptions are frequencies close to $\omega_D$ and $\omega_X$. Near the latter two frequencies, the amplitude $A$ of the second polariton becomes very small.

**K. Goos-Hänchen shift**

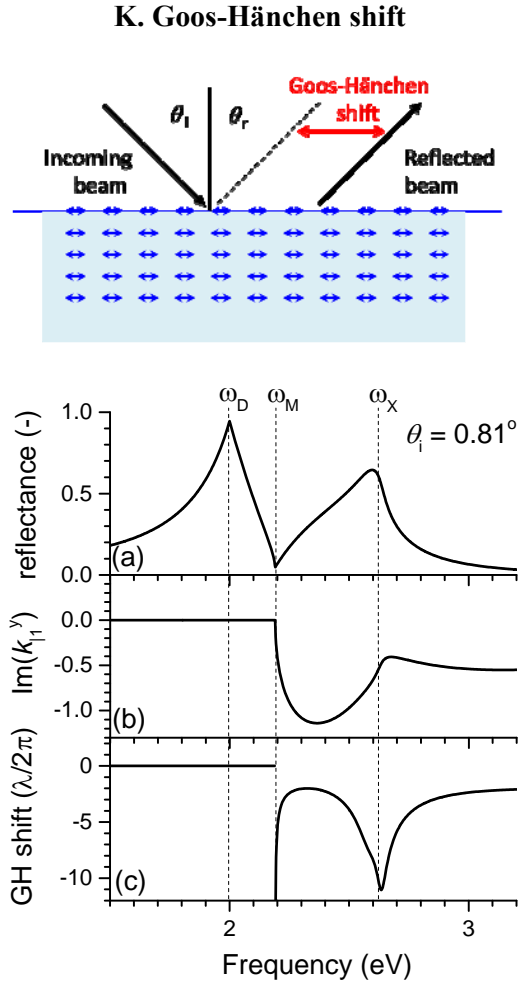

**FIGURE 7**. (Top) Schematic illustration of the Goos-Hänchen (GH) shift. (Lower, a) Reflectance spectrum for light incident along the $xy$ plane with $\theta_i = 0.81°$ and with in-plane polarization. (b) Imaginary part of the $y$-component of the wave-vectors of the refracted beams. (c) Magnitude of the Goos-Hänchen shift in units of the wavelength of the incoming light divided by $2\pi$.

The additional polariton waves discussed in the previous section that travel almost parallel to the matter-vacuum interface. The additional polariton waves may be compared to evanescent waves that occur upon total internal reflection at the dielectric/vacuum interface. Obviously, a major difference between these two types of waves is that the evanescent waves only exist on the vacuum side of the interface, while the polaritons are inside the material. Our reason for bringing up the analogy between the surface type polaritons and the evanescent waves is the following. The total internal reflection process that gives rise to the evanescent wave bound has a remarkable characteristic that is predicted to occur also for the reflection involving polaritons. Upon total internal reflection, the reflected beam is displaced laterally with respect to the incoming beam. This shift is known as the Goos-Hänchen effect, see Fig. 7 for a schematic illustration. This shift was originally demonstrated for total internal reflection at the glass-air interface [58]. The incoming energy can be laterally displaced along the interface through coupling with the evanescent mode on the opposite side of the interface. The Goos-Hänchen shift has also been predicted to occur under specific condition for external reflection at the air-matter interface [59-61]. In this case the incident beam is in the less dense medium in contrast to the case for total internal reflection. The Goos-Hänchen shift has recently been observed experimentally for the reflection from metal surfaces of light incident from the air side of the interface [62]. The magnitude of the Goos-Hänchen shift (GH) can be calculated using the relation

$$GH(\omega, \theta_i) = -\frac{d}{d\theta_i}[\delta(\theta_i)] \tag{73}$$

where $\theta_i$ is the angle of incidence, $\delta(\theta)$ phase shift of the complex reflection. The magnitude of the shift is in units of vacuum wavelength / $2\pi$.

In Fig. 7a, the reflectance near normal incidence is illustrated as a reference. In Fig. 7b we show the imaginary component of the wave-vectors of the associated refracted beams. As can be seen, for frequencies exceeding $\omega_M$, the wave-vectors acquire a negative imaginary component. This implies that both polariton waves are bound to the surface with field amplitude decaying towards zero in the bulk of the material. We note that evanescent waves in an internal reflection process are also characterized by a non-zero imaginary part of the component of the wavevector in the normal direction.

In Fig. 7c we show the magnitude of the Goos-Hänchen shift for reflection of light at the surface of the dipole lattice calculated using (73). We find that calculated shift differs from zero in the frequency range for which also the imaginary part of the $y$-component of the wave-vectors is non-zero. We note that the magnitude of the predicted GH shift is on the order of a wavelength and the theory developed here is therefore be open to experimental falsification.

Based on the coincidence of the GH shift and an imaginary component of the wave vector we provide the following intuitive picture of the reflection process. The incident light first excites a surface bound polariton wave. This is followed by transport of the electromagnetic energy parallel to the surface via the surface bound electromagnetic mode of the medium. Finally release of the surface bound energy back into the vacuum occurs in the form of a reflected wave. We conclude that the minimum in at $\omega_M$ in reflection spectrum is associated with excitation of surface waves. The surface waves gives rise to a Goos-Hänchen shift of the reflected beam.

**L. Comparison to experimental data.**

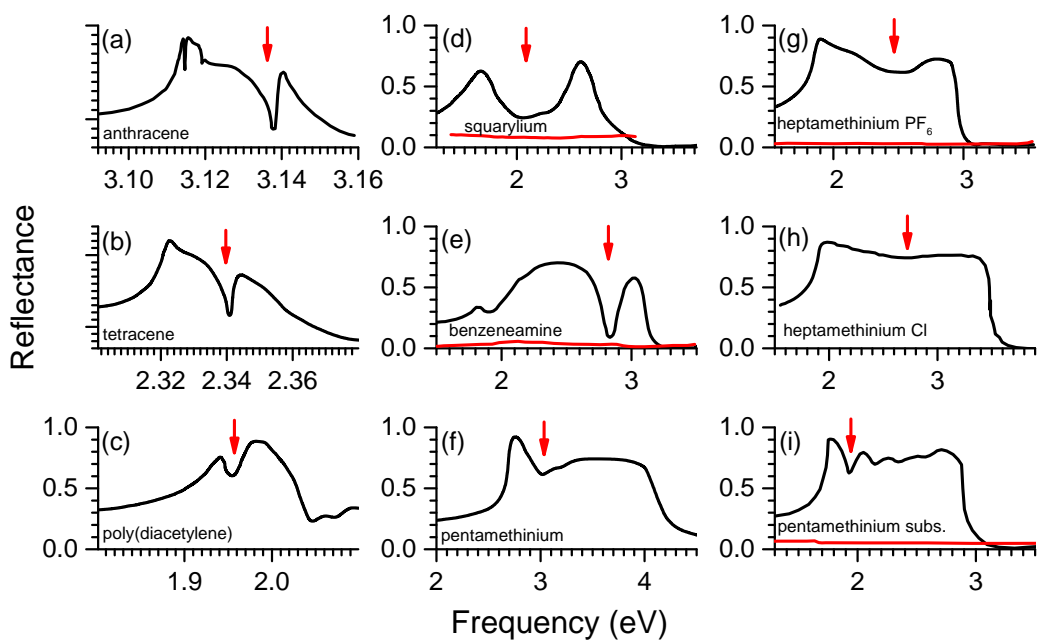

**FIGURE 8**. Experimental data on the reflectance of light from molecular crystals near normal incidence of : (a) anthracene [63,64], (0,0,1) face, pol. // *b*-axis, $T = 2$ K. (b) tetracene [65], (0,0,1) face, pol. // *b*-axis, $T = 2$ K. (c) poly(diacetylene) from bis(*p*-toluenesulfonate)ester of 2,4-hexadiyne-1,6-diol [66], pol. // chains, $T = 4$ K. (d) squarylium dye [67] (2,4-bis-(2-hydroxy-4-dibutylaminophenyl)-1,3-cyclobutadienediylium-1,3-diolate), (1,0,0) face, black and red lines show the reflection polarized along the two principle directions corresponding to maximal and minimal reflectance, room temperature. (e) 4-[4,4-bis[(trifluoromethyl)sulfonyl]-1,3-butadienyl]-N,N-dimethylbenzeneamine [68], (0,1,-1) face, $T = 10$ K. (f) 1,5-bis(dimethylamino)pentamethinium perchlorate [69], (0,1,0) face, $T = 4$ K. (g) 1,7-bis(dimethylamino)hetpmethinium hexafluorophsophate [70], (0,1,0) face, $T = 295$ K. (h) 1,7-bis(dimethylamino)hetpmethinium chloride [71], (0,0,1) face, $T = 295$ K. (i) γ-cyclopropyl-bis(1,3,3-trimethylindolenin-2-yl)pentamethinium tetrafluoroborate [72], (1,1,0) face , $T = 77$ K.

In this last section we compare some of the predictions in the previous section to experimental data. As a first example we discuss the reflectance of anthracene crystals in the frequency range near the lowest allowed singlet transition around 3 eV. In this range, the optical transition from the singlet ground state ($S_0$) to the lowest excited singlet state ($S_1$) occurs. The $S_1 \leftarrow S_0$ transition has a moderate oscillator strength and is polarized in the aromatic plane of the molecule, parallel to the short axis. The 0-0 vibronic band of this transition occurs near 3.1 eV. The (0,0,1) face of anthracene crystals is usually well developed and of sufficient quality for reflection studies. The (0,0,1) facet runs parallel to layers of anthracene molecules stacked in a herringbone arrangement. The long axis of the anthracene molecules in the layers makes an angle of about 30° with the surface normal. The transition dipole moments of the two molecules in the unit cell have large components parallel to the (0,0,1) surface in the direction of the crystallographic *b*-axis. The reflection off the (0,0,1) facet of anthracene has been investigated experimentally in detail [63,73]. The dispersion relation of the polaritons has been investigated [74]. In Fig. 8a, we reproduce the near-normal reflectance recorded at low temperature in the region of the 0-0 transition of the $S_0$-$S_1$ transition. Interestingly, in the low temperature spectra, sharp minima appear. The main minimum occurs near 3.138 eV and is marked by the red arrow in Fig. 8a. Additional minima occur at lower frequency in the 0-0 band. The minima have been assigned to surface waves [75-77]. The occurrence of a minimum in the reflection band is in agreement with the predictions from our model. Moreover, our model affirms the experimentally established relation between the reflectance minimum and the excitation of surface modes, see II K and Fig. 7.

The reflectivity of molecular crystals has been investigated in detail in the past decades and it is well known that for dye molecules with allowed optical transitions strong, metal-like reflection of light can occur at certain crystal faces [78,79]. In Figs 8b-i we have summarized experimental reflection spectra for a number of organic crystals. Spectra involve near-normal reflection on crystals faces with the allowed transition dipole moments of the molecules oriented mainly parallel to the surface. All spectra feature a minimum in the middle of the reflection band associated with the allowed optical transition of the constituent molecules. This is consistent with our analysis of the reflection process involving polaritons, see Fig. 5. We note that optical characterization of oligothiophene crystals [80,81] has yielded experimental evidence for directional dispersion in absorbance spectra that is characteristic of polaritons. The orientation of the molecules with respect to the easily accessible crystal surfaces does however not allow for comparison with predictions based on the model predicted here assuming an orientation of the dipole oscillators parallel to the crystal surface.

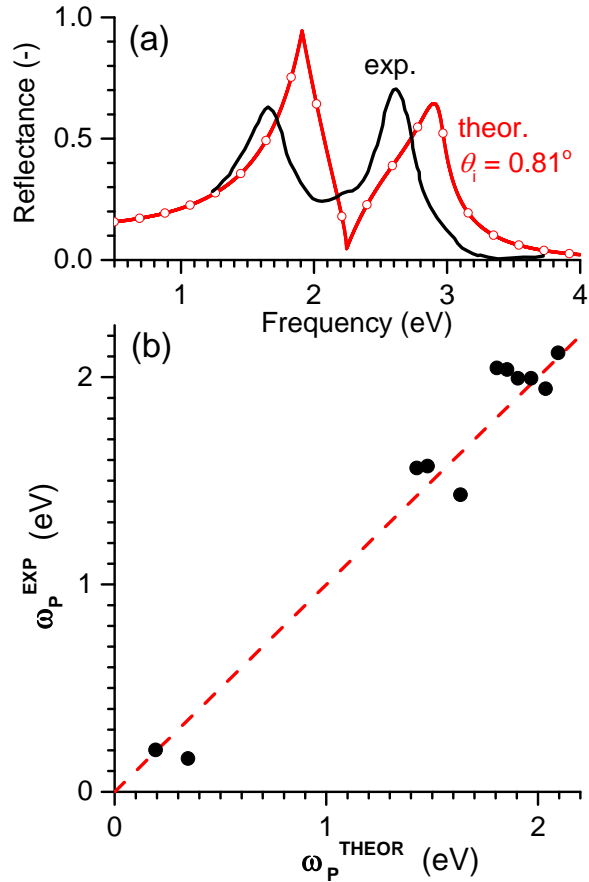

**FIGURE 9**. (a) Experimental reflectance for a crystal of squarylium dye molecules [82] (black line, see also Fig 8d) compared to the reflectance predicted by the dipole oscillator model (red line) for near-normal incidence, assuming a cubic lattice. Apart from the angle of incidence ($\theta_i = 0.81°$) which needs to be near-normal to match the experiment, the model calculation contains *no* other adjustable parameters. $\omega_D = 1.91$ eV and f = 0.947 are taken from absorption of the dye in dilute solution [83] and $\Delta = 4.1 \times 10^{-3}$ eV$^{-1}$ is obtained from the x-ray structure determination [84]. (b) Comparison of values for the plasma frequency $\omega_P$ extracted from experimental reflection measurements (see. Fig. 8) using (75) and values obtained from (74) involving *X*-ray structure determination of the crystal and optical absorption spectroscopy of the dye molecules in solution [30].

In Fig. 9 we compare in detail the reflection spectrum [67,82] for the (1,0,0) face of a crystal of the squarylium dye from Fig. 8d with the prediction of our model. The data for the squarylium dye were chosen because the crystal structure has only one molecule in the unit cell, implying that only one Davydov component needs to be taken into account. The transition dipoles are oriented parallel to the (1,0,0) face. The dye molecule features an intense optical transition in the red part of the spectrum with weak coupling between electronic and nuclear degrees of freedom. A value for the resonance frequency of the isolated dipole oscillator $\omega_D$, can be taken from the position of the allowed transition in the absorption spectrum of a dilute solution of the dye. The oscillator strength $f$ of the lowest allowed transition can be determined from absorption spectroscopy on dilute solutions of the dye and is close to unity. The Thomas-Reiche-Kuhn sum rule [85] states that for a 1-dimensional quantum system, the oscillator strengths of the all the electronic transitions should sum up to the total number of electrons. According to the sum rule, the oscillator strength for a classical dipole oscillator with a moving charge of magnitude $q_e$ should be unity. Therefore the squarylium dye molecule closely resembles a classical dipole oscillator. A value for the lattice spacing $\Delta$ may be obtained by taking the cubic root of the volume of the unit cell. We can then calculate the plasma frequency for the squarylium dye crystal according to:

$$\omega_P^2 = \frac{q^2}{\Delta^3 m_e} = \frac{f 4\pi \alpha_{fine}}{\Delta^3 m_e} \tag{74}$$

with $\alpha_{fine}$ the fine-structure constant. To match the experimental condition of near-normal incidence, we take a value of ~1° for the angle of incidence. We can now calculate the reflection spectrum without any additional adjustable parameters. The resulting spectrum is shown in Fig. 9a. The bandshape of the reflection spectrum can be reproduced quite accurately; agreement between experiment and prediction may be improved further by adjusting the value for $\omega_D$. The separation between the two maxima in the reflection spectrum is determined by the plasma frequency and a value for this variable may be obtained from experiments using the relation:

$$\omega_P^2 = \frac{\omega_X^2 - \omega_D^2}{2} \qquad (75)$$

where $\omega_X$ and $\omega_D$ are frequencies corresponding to the maxima in the reflection. Values for $\omega_P$ have been determined for the reflection spectra shown in Fig. 8. These experimental values values for $\omega_P$ can then be compared to theoretical estimates obtained from relation (74) and using experimentally determined values for $f$ and $\Delta$ (see supporting information for full details). As can be seen, the graph shows a clear correlation between experimental and calculated values.

### III. CONCLUSION

Reflection and refraction of light at the vacuum-matter interface involve a relation between electromagnetic potentials on both sides of the interface. A meaningful relation between potentials can only be given when the two potentials are gauged in the same way. In the case of uniaxial materials, uniformity of gauge across the vacuum-matter interface turns out to be non-trivial because the additional gauge symmetry that exists for massless photons in vacuum is broken for the massive polaritons in bulk matter. Formulation of a Lorentz covariant Lagrange density for the coupled electromagnetic and mechanical degrees of freedom in the uniaxial materials, allows for a new derivation of the boundary conditions for the vector potential at the vacuum-matter interface taking into account the gauge requirement. This approach provides a comprehensive description of optical phenomena at the vacuum-matter interface relating spatial dispersion, surface waves and anomalous refraction. Experimental studies on molecular crystals, metamaterials and/or photonic crystals provide experimental indications for some of the predictions based the model discussed here. Further experimental falsification is required and a unified description of optical phenomena involving metamaterials, molecular, liquid and photonic crystals should be considered.

## ACKNOWLEDGEMENT

We gratefully acknowledge the financial support received from the Dutch Ministry of Education, Culture and Science (Gravity program 024.001.035).